\documentclass[12pt]{article}

\usepackage{amsmath}
\usepackage{amssymb}
\usepackage{mathrsfs}
\usepackage{epsfig}
\usepackage{latexsym}
\usepackage{amssymb}
\usepackage{amsmath}
\usepackage{epsfig}
\usepackage{natbib}
\usepackage[dvips]{color}

\newcommand{\p}{\ensuremath{\mathbb {{}^\prime}}}

\newcommand{\rp}{\ensuremath{\mathbb {RP}}}
\newcommand{\C}{\ensuremath{\mathbb {C}}}
\newcommand{\Z}{\ensuremath{\mathbb {Z}}}

\newcommand{\beq}{\begin{equation}}
\newcommand{\eeq}{\end{equation}}
\newcommand{\bea}{\begin{eqnarray}}
\newcommand{\eea}{\end{eqnarray}}

\numberwithin{equation}{section}

\def\z{\mathfrak{z}}
\def\half{\frac{1}{2}}

\def\Tr{\textrm{Tr}}
\def\H{\textrm{H}}
\def\hsp{,\hspace{.7cm}}
\def\d{\textrm{d}}

\begin{document}

\begin{titlepage}

\setcounter{page}{1} \baselineskip=15.5pt \thispagestyle{empty}

\begin{flushright}
ULB-TH/08-09\\ arXiv:YYMM.NNNNvV
\end{flushright}
\vfil

\begin{center}
{\LARGE Trivializing a Family of Sasaki-Einstein Spaces}
\end{center}
\bigskip\

\begin{center}

{\large Jarah Evslin\footnote{ evslin@sissa.it}}

\end{center}

\begin{center}
\textit{{\it Scuola Internazionale Superiore di Studi Avanzati (SISSA),\\
Strada Costiera, Via Beirut n.2-4, 34013 Trieste, Italia}}
\end{center}

\begin{center}

{\large Stanislav Kuperstein\footnote{skuperst@ulb.ac.be}}

\end{center}

\begin{center}
\textit{{\it Physique Th\'eorique et Math\'ematique,
International Solvay
Institutes, \\ Universit\'e Libre de Bruxelles, ULB Campus Plaine C.P. 
231, 
B--1050 Bruxelles,
Belgium}}
\end{center} \vfil

\noindent 

We construct an explicit diffeomorphism between the Sasaki-Einstein spaces $Y^{p,q}$ and the product space 
$S^3\times S^2$ in the cases $q\leqslant 2$.  When $q=1$ we express the K\"ahler quotient coordinates as an $SU(2)$ 
bundle over $S^2$ which we trivialize.  When $q=2$ the quotient coordinates yield a non-trivial $SO(3)$ bundle 
over $S^2$ with characteristic class $p$, which is rotated to a bundle with characteristic class $1$ and 
re-expressed as $Y^{2,1}$, reducing the problem to the case $q=1$.  When $q>2$ the fiber is a lens space which 
is not a Lie group, and this construction fails.  We relate the $S^2\times S^3$ 
coordinates to those for which the Sasaki-Einstein metric is known.  We check that the RR flux on the $S^3$ is normalized in accordance with Gauss' law and use this normalization to determine the homology classes represented by the calibrated cycles.
As a by-product of our discussion we find a diffeomorphism between $T^{p,q}$ and $Y^{p,q}$ spaces, which means
that $T^{p,q}$ manifolds are also topologically $S^3\times S^2$.

\vfil

\end{titlepage}

\tableofcontents

\pagestyle{headings}


\section{\bf Introduction}

The field theory duals of several infinite families of supersymmetric string 
theory compactifications have been discovered over the last few years
\cite{Gauntlett:2004yd,Martelli:2004wu,Bertolini:2004xf,Benvenuti:2004dy,HEK,Burrington:2005zd,Martelli:2005wy,Cvetic:2005ft,
Benvenuti:2005cz,Benvenuti:2005ja,Franco:2005sm,Cvetic:2005vk,
Butti:2005sw,KMS,EKK}.
The simplest of these families is the set of type IIB string theory compactifications 
on $AdS^5\times Y^{p,q}$ where $Y^{p,q}$ is a 5-dimensional Sasaki-Einstein 
manifold labeled by two integers $p \geqslant q \geqslant 0$. In this paper we will restrict our attention to co-prime 
$p$ and $q$, and so $Y^{p,q}$ is topologically 
(homeomorphic to) the product of a 2-sphere and a 3-sphere \cite{Gauntlett:2004yd}, 
but a set of coordinates for the two spheres is known only in the case 
$Y^{1,0}$ \cite{EK}, which is the base of the conifold.  

Such coordinates would be useful for wrapping branes and for constructing 
orbifolds, but they are difficult to find in general because there is 
not necessarily a calibrated cycle in the homology class of the 3-sphere, 
but only in some multiple of this class which is represented by a lens 
space.  Branes wrapping such non-calibrated cycles may lead to interesting 
effects in the dual gauge theory, where there may be, for example, 
Douglas-Shenker-like \cite{DS} strings or domain walls separating 
discrete sets of vacua.

In the present note we find explicit global coordinates for the 
spheres when $q\leqslant 2$.  This is achieved by using K\"ahler 
quotient coordinates, which are easily transformed into the 
coordinates of the spaces $T^{p,q}$ introduced in \cite{Romans}.  These spaces are quotients of $S^3\times S^3$ by a $U(1)$ which acts on both spheres with weights $p$ and $q$.  Quotienting out by the $U(1)$ action on one of the $S^3$'s turns it into an $S^2$, but a $\Z_q$ subgroup is left unfixed which acts on the other $S^3$.  Quotienting by the remaining $\Z_q$ the other $S^3$ becomes the lens space $L(q;1)$, which is fibered over the $S^2$ with characteristic class $p$.

When $q=1$ the group $\Z_q$ is the trivial group and so the lens space fiber is just $S^3$, which is the group manifold $SU(2)$.  Thus $Y^{p,1}$ is an $SU(2)$ principle bundle over $S^2$, which is necessarily trivial.  We trivialize it.  In the case $q=2$ the lens space is the group manifold $SO(3)$.  Thus $Y^{p,2}$ is an $SO(3)$ principle bundle over $S^2$, with characteristic class $p$.  We may trivialize it on the northern and southern hemispheres and so the bundle is classified by transition functions from the equator to $SO(3)$, or in other words by elements of $\pi_1(SO(3))=\Z_2$.  There are therefore two $SO(3)$ bundles, the trivial and the non-trivial bundle.  As $p$ and $q$ are relatively prime in this note, $p$ is odd, and this implies that our bundle is non-trivial.  $p$ is only a topological invariant modulo 2, and so while we cannot trivialize the bundle, we can rotate it so that its characteristic class becomes 1.  Then we reinterpret $SO(3)$ as an $S^1$ bundle over a new $S^2$ with Chern class 2, which is fibered over the old $S^2$.  Alternately we may consider the $S^1$ to be fibered over the old $S^2$ with characteristic class 1, giving $S^3$, which is then fibered over the new $S^2$ with characteristic class 2.  But this is just $Y^{2,1}$, which we trivialize as before.

In the case $q>2$ this construction fails because the residual $\Z_q$ symmetry is not a normal subgroup\footnote{In this context this point has been emphasized by A. Brini.} of $SU(2)$ and so any change of the characteristic class of the $L(q;1)$ bundle leads to a $\Z_q$ action which is dependent on the position on the $S^2$ base, which mixes the $S^2$ and lens space coordinates and obstructs a reseparation in terms of a different bundle with the $2$-spheres interchanged. 

Ideally one would also like to know the metric in the $S^3\times S^2$ coordinates.  We derive a transformation between coordinates in which the metric is known and our trivialized coordinates in terms of the roots of a certain polynomial.  Numerically it appears as though the solution is indeed unique and so the metric is well-defined.

As an application, we use this construction to tie up a loose end from ~\cite{EKK}.  It was assumed in this work that the calibrated lens spaces $L(j;1)$ represent the $j$th element of the third homology group of $Y^{p,q}$:
\beq
\label{eq:norm}
[L(j;1)]=j\in\H_3(Y^{p,q})=\Z.
\eeq
In fact, the authors found that the known cascade is only reproduced if (\ref{eq:norm}) is true.  The homology class of the lens space determines the overall normalization integrals of the fluxes over the calibrated cycles, and in fact it was implicit in the expressions for the fluxes in \cite{HEK} that (\ref{eq:norm}) holds.  Using our trivialization we obtain a 3-sphere representative of the generator of $\H_3(Y^{p,q})$.  We then explicitly determine the ratio of the normalization of the flux integrated over the generator of the homology, to that of the flux integrated over a calibrated 3-cycle and so confirm (\ref{eq:norm}).   

When $p$ and $q$ are relatively prime, Wang and Ziller \cite{WZ} 
have proven\footnote{We are greatful to James Sparks and Dario Martelli for bringing this paper to our attention.}
that $Y^{p,q}$ is homeomorphic to $S^2\times S^3$.  This proof uses Smale's
classification of simply-connected spin 5-manifolds.  Smale found that such
5-manifolds are completely classified by their second homology group with
integral coefficients.  Using the fact that $Y^{p,q}$ is a circle bundle over
$S^2\times S^2$, the Gysin sequence can be used to find that the second
homology group is just the group $\Z$ of integers, and so Smale's
classification identifies $Y^{p,q}$ as $S^2\times S^3$.  Of course, to use
Smale's classification, one needs to first show that $Y^{p,q}$ is
simply-connected and spin.  In the appendices \ref{Homol} and \ref{Homot} we use the long exact sequence
for homotopy groups of fibrations to show that $Y^{p,q}$ is indeed
simply-connected and we use the Gysin sequence to show that the second homology group is $\Z$.

We begin in Section \ref{trivsec} by finding a homeomorphism between $Y^{p,q}$ and Romans' spaces $T^{p,q}$ and then finding an explicit homeomorphism between these spaces and $S^3\times S^2$ when $q\leqslant 2$.  In Section \ref{metsec} we find the relation between coordinates in which the metric is known and our coordinates, in terms of the roots of a polynomial.  Numerically determining these roots one can then obtain the trivialized metric.  We calculate the RR flux through a representative of the $S^3$ in Section \ref{normsec}, thus establishing (\ref{eq:norm}).  Finally in the appendices 
\ref{Homol} and \ref{Homot}
we discuss the topology of $Y^{p,q}$, obtaining it's homology and homotopy groups.  We find in particular a 1-parameter family of homotopy classes of $S^3\times S^2$ trivializations, corresponding to large diffeomorphisms of $Y^{p,q}$.
We collect some useful $Y^{p,q}$ formulae in \ref{AppC}, while in the last appendix we comment on the $Y^{3,2}$ case.

\section{The construction} \label{trivsec}

\subsection{Notation and conventions}

Throughout this paper we will need a convenient parameterization for the
three- and the two-spheres. The $S^3$ coordinates will be assembled in $2 \times 2$
special unitary  matrices $X \in SU(2)$. 
In this parameterization the ``natural" $\mathbb{R}^4$ coordinates arise through the
Pauli matrix decomposition $X=x_0 \sigma_0 + i \sum_{j=1}^3 x_j \sigma_j$.
Clearly, $\det X=1$ implies that $x_0^2+\sum_{j=1}^3 x_j^2=1$.
As for the $S^2$ there are two possible conventions. One can parametrize the
two-sphere by traceless $SU(2)$ matrices (meaning $x_0=0$), 
which are all anti-hermitian. Alternatively, 
the $S^2$ can be described by the set of $SU(2)$ matrices $S$ subject to the following identification:
\beq
\label{eq:U(1)S}
S \sim S e^{i \lambda \sigma_3},
\eeq
which is nothing but the Hopf projection map.

The former convention was adopted in \cite{EK}, where the $S^2$ matrix was denoted by $Q$.
In this paper, however, we will stick to the latter option. The map between $Q$ and $S$ is given by:
\beq
Q = i S \sigma_3 S^\dagger.
\eeq
Obviously, given $S$ one can find $Q$, which is by construction in $SU(2)$ and traceless,
while starting from $Q$ we can reproduce $S$ exactly up to the $U(1)$ identification
(\ref{eq:U(1)S}).  We will also use the left columns of $X$ and $S$ as the coordinates of the $S^3$ and $S^2$ respectively.

\subsection{The $T^{p,q}$ coordinates on $Y^{p,q}$ and the conifold warm-up example}
\label{eq:Tpqcoord}

The space $Y^{p,q}$ is a $5d$ base of a $6d$ cone, which in turn is a
symplectic reduction of the complex vector space $\C^4$ with weights $\{p,p,-(p-q),-(p+q)\}$.  
In other words the cone over $Y^{p,q}$ is obtained by first solving a $D$-term equation
for the $\C^4$ coordinates $\z_{1,2,3,4}$:
\beq
\label{eq:Dterm}
p \vert \z_1 \vert^2 + p \vert \z_2 \vert^2 - (p-q) \vert \z_3 \vert^2 - (p+q) \vert \z_4 \vert^2 = 0, 
\eeq
which enforces that away from the origin at least one of the first two coordinates, and at least one of the last two coordinates, is non-zero; then one quotients by a $U(1)_K$ action with the above weights. 

It will prove convenient to instead use another set of $\C^4$ coordinates:
\beq
\label{eq:uuvvzzzz}
(u_1,u_{2},v_{1},v_{2}) \propto 
   \left(\z_{1}, \z_{2}, \sqrt{1-\frac{q}{p}} \, \overline{\z}_{3}, \sqrt{1+\frac{q}{p}} \, \overline{\z}_{4} \right).
\eeq
Here we have omitted an overall non-vanishing normalization factor.
The coordinates $u_i$ and $v_i$ still parametrize the $6d$ cone and not the $5d$ base that we are interested in.
Using the $D$-term condition, the $u$ and $v$ two-vectors are non-zero away from the tip and so we may normalize them\footnote{We will address the normalization issue in more detail in Section \ref{metsec}.} to one:
\begin{equation}
\label{eq:length-one}
\vert u_1 \vert^2 + \vert u_2 \vert^2 = \vert v_1 \vert^2 + \vert v_2 \vert^2 =1.
\end{equation}
While both $u_1$ and $u_2$ transform under the $U(1)_K$ with the same weight $p$, the weights of the $v_i$'s are different.  We remedy this by introducing a new two-vector, $w_i$, defined by:
\begin{equation}
\label{eq:w}
\left(
\begin{tabular}{c}
$w_1$\\$w_2$\\
\end{tabular}
\right)=
\left(
\begin{tabular}{cc}
$u_1$&$-\overline{u}_2$\\$u_2$&$\overline{u}_1$\\
\end{tabular}
\right)
\left(
\begin{tabular}{c}
$\overline{v}_1$\\$-v_2$\\
\end{tabular}
\right) ,
\end{equation}
which transforms with weight $q$ and, since the matrix in (\ref{eq:w}) is unitary, it is automatically normalized to length one.  Given $w$ and $u$ one may determine $v$ using (\ref{eq:w}) left-multiplied by the inverse of the
$u$-matrix.  Therefore, for fixed $u$, (\ref{eq:w}) provides a one-to-one map between values of $v$ and $w$.  While $(u,v)$ is a pair of symplectic quotient coordinates for $Y^{p,q}$, $(u,w)$ is a pair of 3-spheres with a common $U(1)_K$ action with weights $p$ and $q$ respectively, identifying it as a set of coordinates for the space $T^{p,q}$ of \cite{Romans}, although the metric is not the same.

For later use we will introduce the following matrices:
\beq
\label{eq:UVW}
U \equiv \left(
\begin{tabular}{cc}
$u_1$&$-\overline{u}_2$\\$u_2$&$\overline{u}_1$\\
\end{tabular}
\right), \qquad
V \equiv \left(
\begin{tabular}{cc}
$v_1$&$-\overline{v}_2$\\$v_2$&$\overline{v}_1$\\
\end{tabular}
\right) \qquad \textrm{and} \qquad
W \equiv \left(
\begin{tabular}{cc}
$w_1$&$-\overline{w}_2$\\$w_2$&$\overline{w}_1$\\
\end{tabular}
\right),
\eeq
which transform under the $U(1)_K$ as:
\beq
\label{eq:UVW-K}
U \to U e^{i p \lambda \sigma_3}, \qquad
V \to e^{-i q \lambda \sigma_3} V e^{i p \lambda \sigma_3}, \qquad \textrm{and} \qquad
W \to W e^{i q \lambda \sigma_3}.
\eeq
Now (\ref{eq:w}) reduces to
\beq
\label{eq:WUV}
W\equiv U V^\dagger.
\eeq
In summary, we have demonstrated that the spaces $T^{p,q}$
and $Y^{p,q}$ are homeomorphic and the explicit map is given by (\ref{eq:w}).
In  particular, $T^{1,0}$ has the same topology as $Y^{1,0}$.
The former is topologically $S^3 \times S^2$ since $U$ in (\ref{eq:UVW-K}) is $U(1)_K$-invariant
and therefore parameterizes a three-sphere, while $W$ transforms with weight one and 
so, like $S$ in (\ref{eq:U(1)S}), describes a two-sphere via the Hopf map.
As for $Y^{1,0}$, it has precisely the conifold base charges $\{1,1,-1,-1\}$.
Thus we learn that:

\beq
T^{1,1} \equiv Y^{1,0} \cong T^{1,0} \equiv S^3 \times S^2.
\eeq
Let us end this section by showing that the conifold trivialization obtained here 
coincides with the result of \cite{EK}.
We found that the three-sphere is given by $X=W$, while for the two-sphere
we can choose between $S=U$ and $S=V$.  In what follows we will prefer the latter option.
By definition $X$ satisfies:
\beq
u=Xv,
\qquad \textrm{where} \qquad
u \equiv \left(
\begin{tabular}{c}
$u_1$\\$u_2$\\
\end{tabular}
\right)
\quad \textrm{and} \quad
v \equiv \left(
\begin{tabular}{c}
$v_1$\\$v_2$\\
\end{tabular}
\right) . 
\eeq
As $X \in SU(2)$ there is a unique solution for $X$ in terms of the vectors $u$ and $v$:
\beq
\label{eq:uXv}
X = u v^\dagger - \epsilon \overline{u} v^{\textrm{T}} \epsilon = 
  u v^\dagger - \left( \overline{u} v^{\textrm{T}} \right)^{\textrm{T}} + \Tr (\overline{u} v^{\textrm{T}}) \cdot \sigma_0.
\eeq
The conifold is defined by a complex $2 \times 2$ singular matrix $\Omega$ (or $W$ in the ``standard"
notation used in \cite{EK}). To properly describe the conifold base $T^{1,1}$, the matrix $\Omega$ has to be normalized as 
$\Tr(\Omega^\dagger \Omega)=1$. Furthermore, in terms of $u$ and $v$ we have $\Omega=u v^\dagger$, which 
obviously renders
$\Omega$ invariant under the $U(1)_K$ quotient. Finally, substituting this into (\ref{eq:uXv}) we arrive at
\beq
X = \Omega - \Omega^\dagger + \left( \Tr \Omega^\dagger \right) \cdot \sigma_0,
\eeq
which is exactly the $S^3$-projection proposed in \cite{EK}.  The inverse map, of course, is given by 
$\Omega=X v v^\dagger$,  where $v$, in turn, is fixed by $V=S$.

\subsection{Trivializing $Y^{p,1}$}

In this section we will construct a homeomorphism between $Y^{p,1}$ and $S^3\times S^2$.
For $q=1$ the weights of $U$ and $W$ in (\ref{eq:UVW-K}) are $p$ and $1$ respectively,
so we can use $W$ to parameterize the $S^2$, since it transforms exactly like $S$ in (\ref{eq:U(1)S}).
Next let us define a weight $p$ unitary matrix:
\begin{equation}
\label{eq:hatW}
\widehat{W}=
c_{\widehat{W}}
\left(
\begin{array}{cc}
w_1^p & -\overline{w}_2^p \\
w_2^p & \overline{w}_1^p 
\end{array}
\right) ,
\end{equation}
where $c_{\widehat{W}}$ is a normalization constant $(\sqrt{|w_1|^{2p}+|w_2|^{2p}})^{-1}$, which normalizes $\widehat{W}$ to have determinant one.  The matrix $\widehat{W}$ transforms with weight $p$, namely 
$\widehat{W} \to \widehat{W} e^{i p \lambda \sigma_3}$. 
Now the $S^3 \times S^2$ parameterization simply reads:
\beq
\label{eq:XSq1}
X = U \widehat{W}^\dagger
\qquad \textrm{and} \qquad
S = W .
\eeq
Clearly, $X \in SU(2)$ and is $U(1)_K$-invariant. Thus $X$ is a good coordinate for $S^3$ and $S/U(1)_K$ is a good coordinate for $S^2$.  Moreover, the map is invertible. Indeed, given $X$ and any representative of $S$ we can compute $\widehat{W}$ using $W=S$ and then find $U$ using $U=X \widehat{W}$.

\subsection{Trivializing $Y^{p,2}$}

To trivialize $Y^{p,2}$ we will again use the $(u_1,u_2)$ and the $(w_1,w_2)$
coordinates for $T^{p,q}$ introduced in (\ref{eq:w}). The former have weight $p$ and the latter
weight $q$ under the $U(1)_K$ action (\ref{eq:UVW-K}).  The trivialization will occur in five steps.

\begin{itemize}
\item[\emph{1}.] First we begin with $Y^{p,2}$ described as a K\"ahler quotient.  As we have seen, the solutions of the $D$-term condition yield an $S^3\times S^3$ whose quotient by $U(1)_K$ is $Y^{p,2}$:
\begin{equation}
\begin{array}{ccc} 
U(1)_K & \longrightarrow & S^3 \times S^3 = \{(u,v)\}             \\
       &                 &  \bigg\downarrow  \scriptstyle{c_1=1}  \\
       &                 &  Y^{p,2}   \phantom{xxx}
\end{array}
\end{equation}
\item[\emph{2}.] Then we use (\ref{eq:w}) to pass to $T^{p,2}$ coordinates $(u,w)$.  We still have a quotient of $S^3\times S^3$ by $U(1)_K$:
\begin{equation}
\begin{array}{ccc} 
U(1)_K & \longrightarrow & S^3 \times S^3 = \{(u,w)\}             \\
       &                 &  \bigg\downarrow  \scriptstyle{c_1=1}  \\
       &                 &  T^{p,2}   \phantom{xxx}
\end{array}
\end{equation}
\item[\emph{3}.] We quotient $w$ by the $U(1)_K$ action, leaving an $S^2$.  A $\Z_2$ subgroup of the $U(1)_K$ is not fixed by this gauge choice for $w$.  This $\Z_2$ acts on the $u$ coordinate yielding $SO(3)=S^3/\Z_2$.  The $SO(3)$ is fibered over the $S^2$ with characteristic class $p=2k+1$:
\begin{equation}
\begin{array}{ccc} 
\{u\}/\Z_2=SO(3)     & \longrightarrow & T^{p,2} = \{(u,w)\}/U(1)_K                    \\
                     &                 &   \bigg\downarrow  \scriptstyle{c_1=2k+1}     \\
                     &                 &  S^2 = \{w\}/U(1)_K  \phantom{x}
\end{array}
\end{equation}
\item[\emph{4}.] The characteristic class $c_1$ of the $SO(3)$ fibration is only a topological invariant modulo 2.  Therefore we may change the coordinates so that it decreases from $2k+1$ to $\pm 1$.  Now we have a circle fibered over the $u^\prime$
two-sphere with Chern class 2, which is in turn fibered over the $w$ two-sphere with Chern class 1:
\begin{equation}
\begin{array}{ccc} 
\{u^\prime\}/\Z_2=SO(3)         & \longrightarrow & T^{p,2} = \{(u^\prime,w)\}/U(1)_K           \\
                                &                 &  \bigg\downarrow  \scriptstyle{c_1=\pm 1}   \\
                                &                 &  S^2 =  \{w\}/U(1)_K  
\end{array}
\end{equation}

\item[\emph{5}.] We switch the orders of the two $S^2$'s, so that $S^1$ is fibered over $S^2$ with Chern class 1, yielding an $S^3$, which is fibered over the other $S^2$ with Chern class 2.  This is just $Y^{2,1}$, which we may trivialize as in the previous subsection.  In practice this switch occurs by introducing a new gauge degree of freedom $U(1)^\prime_K$ and then choosing a gauge for the $u^\prime$ coordinate which entirely fixes the gauge symmetry.  Thus in the end the gauge-fixed $u^\prime$ is quotiented to $S^2$ and the new $w$ is gauge-independent, and so parameterizes the $S^3$:
\begin{equation}
\begin{array}{ccc} 
\{w^\prime\}=S^3     & \longrightarrow & Y^{2,1} = \{(u^{\prime\prime},w^\prime)\}/U(1)^\prime_K   \\
                     &                 &   \bigg\downarrow  \scriptstyle{c_1=2}                    \\
                     &                 &  S^2 =  \{u^{\prime\prime}\}/U(1)^\prime_K  
\end{array}
\end{equation}

\end{itemize}

As the first two steps have been performed in the previous subsections, we will begin with the third step.

The points of $Y^{p,q}$ correspond to orbits of the $U(1)_K$ action.
We can obtain $Y^{p,q}$ coordinates by fixing the gauge.
Let us denote the phases of $w_1$ and $w_2$ by $\psi_1$ and $\psi_2$
respectively.
One convenient gauge choice is $\psi_1=0$, which, in turn,
corresponds to $\lambda_N=-\psi_1/q$ in (\ref{eq:UVW-K}).
This gauge choice, however, is not defined on all of $Y^{p,q}$, because when $w_1=0$, $\psi_1$ is not well-defined.

The $(w_1,w_2)$ coordinates alone, quotiented by the $U(1)_K$ action, 
define the Bloch sphere $S^2$ with north pole $w_2=0$ and south pole $w_1=0$.  
We can cover the $S^2$ by two open discs, the northern patch $S^2_N$ in which $w_1\neq 0$ and the southern patch $S^2_S$ in which $w_2\neq 0$.  Then the gauge choice $\psi_1=0$ is well defined on $S^2_N$.  On $S^2_S$ one may choose the gauge condition $\psi_2=0$ or, equivalently, $\lambda_S=-\psi_2/q$.
Summarising, on the northern patch $S^2_N$ we fixed the gauge by setting $\psi_1=0$,
while on the southern patch  $S^2_S$ we have $\psi_2=0$. Notice that none of the above choices fixes the gauge completely. Instead we have a residual discrete transformation that acts as 
$(u_1,u_2) \to (\eta_q u_1,\eta_q u_2)$, where $\eta_q$ is the $q$th root of unity.
We learn, therefore, that $u_1$ and $u_2$ become the coordinates for the lens space $L(q;1)$.
Therefore $w/U(1)_K$ parameterizes an $S^2$ and at each point on $S^2$, $u/\Z_q$ parameterizes an $L(q;1)$.  A similar argument was used in \cite{Martelli:2004wu} to show that for $u_{1,2}=0$, $v_1=0$ and 
$v_2=0$ one finds $L(p;p-1)$, $L(p+q;1)$ and $L(p-q;1)$ respectively. 

Gluing the patches $S^2_N$ and $S^2_S$ together one obtains $Y^{p,q}$ described as an $L(q;1)$ bundle 
over $S^2$ with local trivializations $\psi_1=0$ and $\psi_2=0$. 
The characteristic class of this bundle is given by the winding number of the transition function that 
relates $u$'s in the northern patch to $u$'s in the southern patch.
This transition function is equal to the ratio of the two values of $u$'s, which is 
$e^{i p (\lambda_N - \lambda_S)}=e^{i\frac{p}{q}(\psi_2-\psi_1)}$.  
This is a map from the overlap of the two patches to a $U(1)_K$ in the structure group of the bundle.  
As one goes around the overlap once, say by going once around the $S^2$ equator, $\psi_2-\psi_1$ increases 
by one unit.
The transition function then increases in phase by $2\pi\frac{p}{q}$.
An increase by $2\pi/q$ takes a point on $L(q;1)$ to itself, as we have quotiented out by $q$th roots of unity.
Thus the smallest well-defined transition function, corresponding to characteristic class
equal to one, would increase in phase by $2\pi/q$ as one circumnavigates the equator.  The current transition
function has a winding number which is $p$ times higher, and so it corresponds to an $L(q;1)$ bundle over
$S^2$ with characteristic class equal to $p$.

Now we will restrict our attention to the case $q=2$ in which:
\beq
L(2;1)=\rp^3=SO(3).
\eeq
The parameters $p$ and $q$ are taken to be relatively prime, and so $p$ is odd and we may write:
\beq
p=2k+1
\eeq
for some positive integer $k$.  The bundle is principle and so the transition functions are maps from the equatorial $S^1$ to the structure group $SO(3)$.  Therefore the bundles are classified by a topological invariant with values in $\pi_1(SO(3))=\Z_2$ and so the characteristic class is only invariant modulo 2.  This implies that there exists some rotation with winding number $-k$ which will shift the characteristic class by $-2k$ so that it decreases from $2k+1$ to $1$.  On the northern patch this rotation must be well-defined everywhere and in particular at $w_2=0$, but at $w_1=0$, which is not part of the patch, it should change the winding number of $u$ with respect to the $S^2$ equatorial coordinate $\psi_2-\psi_1$ by $\frac{2\pi}{q} \cdot 2k$, so that the transition function shifts by $2k$ units.  Similarly on the southern patch it must be well-defined at $w_1=0$ but shift the winding number at the north pole by $2k$ units.

One such rotation is:
\begin{equation}
\label{eq:RotationA}
\left(
\begin{tabular}{c}
$u_1^\prime$ \\ $\overline{u}_2^\prime$ \\
\end{tabular}
\right)
=
c_{\widetilde{W}}
\left(
\begin{tabular}{cc}
$\overline{w}^k_1$ & $w_2^k$ \\ $-\overline{w}_2^k$ & $w_1^k$\\
\end{tabular}
\right)
\left(
\begin{tabular}{c}
$u_1$\\ $\overline{u}_2$ \\
\end{tabular}
\right)
\qquad
\textrm{with}
\quad 
c_{\widetilde{W}} = (\sqrt{|w_1|^{2k}+|w_2|^{2k}})^{-1}
\end{equation}
\beq
\label{eq:RotationB}
\textrm{or} \qquad u_1^\prime=c_{\widetilde{W}}( \overline{w}^k_1 u_1 + w_2^k \overline{u}_2)
\qquad \textrm{and} \qquad
\overline{u}_2^\prime= c_{\widetilde{W}}( -\overline{w}^k_2 u_1 + w_1^k \overline{u}_2).
\eeq
We will verify first that the rotation is well defined. 
On the northern patch ($w_1 \neq 0$ and $\lambda_N=-\psi_1/2$) the first terms on the right hand side  of the two expressions
in (\ref{eq:RotationB})
have charge one under $U(1)_K$ and so are multiplied by the $e^{-i\psi_1/2}$ factor, while the second terms have instead
$e^{i\psi_1/2}$.  The same observations hold on the southern patch with $\psi_1$
replaced by $\psi_2$.
When $\psi_{1,2}\rightarrow\psi_{1,2}+2\pi$ both terms change sign,  
so the whole expression still defines the same element of $\rp^3$. 
This would not have been the case for $q>2$ because the two terms in $u^\prime_{1,2}$ would 
have changed by different weights ($e^{-i\psi_{1,2}/q}$ and $e^{i\psi_{1,2}/q}$), implying that the $\Z_q$ 
identification of the $S^3$ 
would have depended on the $w$ coordinate.  
This is a reflection of the fact that $\Z_q$ is a normal subgroup of $SU(2)$ only for $q \leqslant 2$.

Let us now show that (\ref{eq:RotationA},\ref{eq:RotationB}) indeed shifts the transition functions by $2k$ units.
Unfortunately, the north and south $(u^\prime_1,u^\prime_2)$ vectors are no longer proportional, 
and so the transition function in general is quite complicated.  
However, the dependence greatly simplifies near the poles.
To be more specific, $u_{1,2}^\prime \approx u_{1,2} \overline{w}_1^k$ near the north pole
and $u_1^\prime \approx \overline{u}_2 w_2^k$, $u_2^\prime \approx - \overline{u}_1 w_2^k$  
near the south pole.
Therefore, one may calculate the transition functions near the poles to evaluate the 
characteristic class of the bundle. 
Comparing the ratio $u_1^\prime/u_2^\prime$ near the north pole for $\lambda=\lambda_N$ and
$\lambda=\lambda_S$ we find that the transition function is $e^{(\psi_2-\psi_1)/2}$.
Similarly near the south pole we have $e^{(\psi_1-\psi_2)/2}$.
Thus as one encircles the $S^2$ once, the transition function has winding number $1$ with respect 
to $\rp^3$ near the north 
pole, and $-1$ near the south pole.

Such position-dependent characteristic classes are to be expected, 
as $\rp^3$ is $SO(3)$ and the transition functions of $SO(3)$ are valued in $\pi_1(SO(3))=\Z_2$, 
and so all odd numbers are equivalent.  However, if we insist on fixing an integral characteristic class we can.  
For example, one may consider the southern patch to be just a small neighbourhood of the south pole.  
Then the transition function may be slightly deformed to be just multiplication by the phase $e^{(\psi_1-\psi_2)/2}$ 
and so the characteristic class is equal to $-1$.  
If instead one made the northern patch small, one would conclude that the characteristic class is $1$.  
Fortunately, our final construction will be globally well-defined and so no such choice will be necessary 
in the end.

Summarising, the coordinates $u\p_{1,2}$ describe an $\rp^3$ fibered over the $S^2$ parameterized by the gauge fixed 
$w_{1,2}$.  This fibration now has characteristic class $\pm 1$.  $\rp^3$ is a circle bundle over $S^2$ 
with Chern class equal to 2, so $S^1$ is fibered over an $S^2$ with Chern class 2 which is all 
together fibered over another $S^2$ with Chern class 1.  Our goal is to interchange these two Chern classes, 
because we will then obtain $Y^{2,1}$, which, as we have showed above, is\footnote{Recall that $L(1;1)=S^3$.}
an $S^3$ fibered over $S^2$ with Chern class 2.
We know how to trivialize $Y^{2,1}$ and so then we will be done.

To interchange the two Chern classes we will migrate the circle fiber from the $u=(u_1,u_2)$ two-sphere to the 
$w=(w_1,w_2)$ two-sphere.
The $S^1$ originally was fibered over both, before we fixed the gauge.  The circle is fibered over $u$ and not $w$ 
because we fixed the gauge by fixing a phase in $w$ (more precisely, we fixed the phases of $w_1$ and $w_2$
on the northern and the southern patches respectively).
Had we instead fixed a phase in $u$ then the circle would have 
been fibered over $w$.  Therefore our strategy will be to re-introduce a new $U(1)^\prime_{K}$ gauge freedom, 
so that the circle is 
again fibered over both spheres, yielding the K\"ahler coordinates for $Y^{2,1}$, and then we will fix this new gauge 
freedom by fixing the phase of $u\p$ so that the circle is fibered over $w$, yielding an $S^3$ which is fibered 
over the $u\p$ two-sphere with characteristic class 2.

Since our construction necessarily involves gauge fixing the original $U(1)_K$ symmetry,
we have to consider separately the northern and the southern patches. We will see, however, that the final result
is globally well-defined, so focusing only on one of the two patches is sufficient.
We will choose the southern patch, $w_2 \neq 0$. The gauge fixing $\lambda_S=-\psi_2/2$ can be recast 
in the following form:
\beq
u_{1,2}^{\textrm{f}} = u_{1,2} e^{-i (k+\frac{1}{2}) \psi_2} = u_{1,2} \left( \frac{\overline{w}_2}{\vert w_2 \vert} \right)^{k+\frac{1}{2}},
\quad
w_1^{\textrm{f}} = w_1 e^{-i \psi_2} = w_1 \frac{\overline{w}_2}{\vert w_2 \vert},
\quad 
w_2^{\textrm{f}} = w_2 e^{-i \psi_2} = \vert w_2 \vert.
\eeq
Next we substitute these gauge fixed $L(q;1) \times S^2_N$ coordinates $u^{\textrm{f}}$ and $w^{\textrm{f}}$
into (\ref{eq:RotationB}):
\beq
\label{eq:RotationC}
u_1^\prime=c_{\widetilde{W}}
 \left(  u_1 \overline{w}^k_1 \frac{\overline{w}_2^{1/2}}{\vert w_2 \vert^{1/2}} 
                              + \overline{u}_2 \frac{w_2^{k+1/2}}{\vert w_2 \vert^{1/2}}  \right)
\qquad \textrm{and} \qquad
u_2^\prime= c_{\widetilde{W}} 
 \left(  u_2 \overline{w}^k_1 \frac{\overline{w}_2^{1/2}}{\vert w_2 \vert^{1/2}} 
                              - \overline{u}_1 \frac{w_2^{k+1/2}}{\vert w_2 \vert^{1/2}}  \right).
\eeq
Now we have to introduce a new gauge $U(1)^\prime_K$. It acts as:
\beq
u_{1,2}^{\prime \prime} = u_{1,2}^\prime e^{\frac{i}{2} \psi_2} 
      = u_{1,2}^\prime \left( \frac{w_2}{\vert w_2 \vert} \right)^{1/2},
\quad
w_1^{\prime} = w_1^{\textrm{f}} e^{i \psi_2} = w_1,
\quad 
w_2^{\prime} = w_2^{\textrm{f}} e^{i \psi_2} = w_2.
\eeq 
Notice that the new gauge has a $\Z_2$ ambiguity, which cures the same ambiguity in the $U(1)_K$
gauge fixing. This miracle would have failed for $q>2$.  Moreover, $U(1)^\prime_K$ reproduces the original $w$ coordinates.
As for $u_1^{\prime \prime}$ and $u_2^{\prime \prime}$, we find from (\ref{eq:RotationC}) that:
\beq
\label{eq:RotationD}
u_1^{\prime \prime} = c_{\widetilde{W}}
 \left(  u_1 \overline{w}^k_1 + \overline{u}_2 \frac{w_2^{k+1}}{\vert w_2 \vert}  \right)
\qquad \textrm{and} \qquad
u_2^{\prime \prime} = c_{\widetilde{W}} 
 \left(  u_2 \overline{w}^k_1 - \overline{u}_1 \frac{w_2^{k+1}}{\vert w_2 \vert}  \right).
\eeq
As advertised, the new coordinates $w^\prime$ and $u^{\prime \prime}$  have weights $2$ and $1$
with respect to the new gauge $U(1)^\prime_K$, so we have successfully arrived at a set of $Y^{2,1}$
coordinates. Moreover, the final expressions (\ref{eq:RotationD}) are well-defined also at the north pole, 
where $w_2=0$. Indeed, since $k \geqslant 1$ the limit $w_2 \to 0$ of (\ref{eq:RotationD}) is absolutely smooth.

Using the results of the previous subsection it is straightforward to find the connection between the $S^3 \times S^2$
coordinates and the original $u$ and $w$ coordinates on $Y^{p,2}$.
The result can be presented in a simple form like in the $q=1$ case, if we will define the following matrix:
\begin{equation}
\widetilde{W} = 
c_{\widetilde{W}}
\left(
\begin{array}{cc}
w^k_1 & {\displaystyle - \frac{\overline{w}_2^{k+1}}{|w_2|} } \\ {\displaystyle\frac{w_2^{k+1}}{|w_2|}} & \overline{w}_1^k\\
\end{array}
\right) .
\end{equation}
The matrix transforms under $U(1)_K$ as: 
$\widetilde{W} \to e^{-i \lambda \sigma_3}\widetilde{W} e^{i p \lambda \sigma_3}$.
With this definition the $S^2$ matrix $S$ is given by:
\beq
\label{eq:SUW}
S = U \widetilde{W}^\dagger,
\eeq
where $U$ is defined as in (\ref{eq:UVW}) and transforms as in (\ref{eq:UVW-K}). 
One can easily check that $U(1)_K$ acts on $S$ as in (\ref{eq:U(1)S}). 
Since the $u^{\prime\prime}$ coordinates now play the r\^ole of the $S^2$ coordinates $s$, we have to 
define a new matrix $\widehat{S}$ the same way (\ref{eq:hatW}) that we defined $\widehat{W}$ in the $(p,q)=(2,1)$ case:
\begin{equation}
\widehat{S}=
c_{\widehat{S}}
\left(
\begin{array}{cc}
s_1^2 & -\overline{s}_2^2 \\
s_2^2 & \overline{s}_1^2 
\end{array}
\right) ,
\quad \textrm{with} \quad c_{\widehat{S}}=(\sqrt{|s_1|^{4}+|s_2|^{4}})^{-1}.
\end{equation}
Since $\widehat{S}$ transforms as $\widehat{S} \to \widehat{S} e^{2 i \lambda \sigma_3}$,
the $S^3$ matrix
\beq
\label{eq:XSW}
X = \widehat{S} W^\dagger
\eeq
is gauge invariant.  Thus $S$ and $X$ are global coordinates for $S^2$ and $S^3$ respectively.

To summarise, starting from the $u$ and $w$ coordinates on $Y^{p,2}$ we may find the $S^2$ coordinate S from (\ref{eq:SUW}). These coordinates can be further used to find the $S^3$ coordinate $X$ using (\ref{eq:XSW}).
The inverse map is given by:
\beq
W = X^\dagger  \widehat{S}
\qquad \textrm{and} \qquad
U = S \widetilde{W},
\eeq
where, again, the first formula has to be substituted into the second.

\section{Trivialized coordinates in terms of metric coordinates} \label{metsec}
\label{TrivCoord}

In this section we would like to find a relation between the coordinates of the $5d$ $Y^{p,q}$ metric and 
the explicit $S^3 \times S^2$ coordinates identified in the previous section. 
To achieve this goal we first have to 
re-write the $\C^4$ coordinates $\z_1$, $\z_2$,
$\z_3$, and $\z_4$ in terms of the metric coordinates.
We will pursue the following strategy. The cone over $Y^{p,q}$ can be alternatively
defined as a set of all possible $U(1)_K$-invariant $\z_i$ monomials quotiented by all
possible relations among them. There are three families of $2p+5$ independent monomials in this algebra:
\begin{eqnarray}
\label{eq:abc}
&&  a_i = \z_1^j \z_2^{(p-q)-j} \z_3^p \quad \textrm{with} \quad j=0, \ldots p-q    \nonumber \\
&&  b_0 = \z_1^2 \z_3 \z_4, \quad b_1 = \z_1 \z_2 \z_3 \z_4, \quad b_2 = \z_2^2 \z_3 \z_4 \nonumber \\
&&  c_i = \z_1^j \z_2^{(p+q)-j} \z_4^p \quad \textrm{with} \quad j=0, \ldots p+q .
\end{eqnarray}
On the dual gauge theory side these variables correspond to the gauge invariant mesonic operators
(see \cite{Benvenuti:2004dy} and \cite{KMS} for a detailed description).
From a geometric point of view, these are regular (holomorphic) solutions of the $6d$
Laplacian equation. Using (\ref{eq:abc}) we can, therefore, express $\z_i$'s
in terms of the metric coordinates. These expressions, of course,  will necessarily include
a free complex parameter. The absolute value of this parameter has to be fixed by the $D$-term equation
(\ref{eq:Dterm}), 
while the phase corresponds to the $U(1)_K$ gauge freedom.

The $6d$ cone metric is $d s^2_{(6)} = d r^2 + r^2 d s^2_{(5)}$ and
the $5d$ metric on $Y^{p,q}$ is given by:
\begin{eqnarray}
\label{eq:5d}
   \d s^2_{(5)} &=& \frac{1-y}{6} \left(\d \theta^2 + \sin^2\theta \d \phi^2 \right) + 
        \frac{\d y^2}{w(y)q(y)}  + \frac{q(y)}{9} \left( \d\psi - \cos\theta \d \phi \right)^2  + \nonumber \\   
     && + {w(y)} \left( \d \alpha +\frac{a-2y+y^2}{6(a-y^2)}
         \left( \d\psi-\cos\theta \d\phi \right) \right)^2
\end{eqnarray}
with
\label{eq:w(y)q(y)}
\beq
 w(y) = \frac{2(a-y^2)}{1-y}
\qquad \textrm{and} \qquad
 q(y) = \frac{a-3y^2+2 y^3}{a-y^2}~.
\eeq
Both $\phi$ and $\psi$ are $2\pi$ periodic, while the coordinates $\theta$ and $y$ span the range:
\beq
0 \leqslant \theta \leqslant \pi
\qquad \textrm{and} \qquad y_1 \leqslant y \leqslant y_2,
\eeq
where the constants $y_1$ and $y_2$ are the smallest two roots of the numerator of $q(y)$
in (\ref{eq:w(y)q(y)}) and are determined by:
\beq
y_{1,2} = \frac{1}{4 p} \left( 2 p \mp 3 q - \sqrt{4p^2-3q^2} \right).
\eeq
These relations also fix the constant $a$ in (\ref{eq:w(y)q(y)}). In what follows we will denote the biggest root 
of the numerator by $y_3$.
Finally, the period of $\alpha$ according to the literature is $2 \pi \ell$, where:
\beq
\ell \equiv \frac{q}{3 q^2 -2p^2+p\sqrt{4p^2-3q^2}}.
\eeq
This result first appeared in \cite{Gauntlett:2004yd}. It was argued there that the space parameterized by the coordinates
$\theta$, $\phi$, $y$ and $\psi$ is topologically $S^2 \times S^2$, while $\alpha$ describes a circle
fiber over this base. To avoid singularities the periods of the $U(1)$-connection over the two spheres
should satisfy $P_1/P_2=p/q$, where $p$ and $q$ are two co-prime integers \cite{Gauntlett:2004yd}. 
A straightforward calculation
then produces the above result for $\alpha$.
We find a similar, but not identical result. The points $\theta=0,\pi$
correspond to $\z_2=0$ and $\z_1=0$ respectively. On the other hand, we know that at $\z_{1,2}=0$
the space reduces to the lens space $L(p;p-1)$. So we might check directly whether for these values of $\theta$
the periods of $\phi$, $\psi$ and $\alpha$ match those of the lens space. A similar check can be performed for $y=y_{1,2}$. 
A direct calculation reveals that the angles $\phi$ and $\psi$ are $2 \pi$-periodic, but the third  angle 
with this period should be:
\beq
\label{eq:alphaNEW}
\tau = \frac{p+q}{2}(\phi+\psi) + \frac{\alpha}{\ell}
\eeq
and not $\alpha/\ell$ alone as was advocated in \cite{Gauntlett:2004yd}.
Remarkably, our result does not differ from \cite{Gauntlett:2004yd} when $p+q$ is even. Unfortunately,
we don't know the origin of this discrepancy.
We will come back to this point later in this section.

The $6d$ Laplace equation $\Box_{(6)} Z =0$ has three independent solutions \cite{Berenstein:2005xa,Burrington:2005zd}:
\beq
Z_1 = \tan \frac{\theta}{2} \, \, e^{\textstyle{i \phi}}, \quad
Z_2 = \frac{1}{2} \sin \theta \, \, 
               e^{\textstyle{- 6 \int \! \frac{\d y}{w(y)q(y)} + i (6 \alpha + \psi)}}, \quad
Z_3 = \frac{r^3}{2} \sin \theta \, \, e^{\textstyle{- 6 \int \! \frac{y\d y}{w(y)q(y)} + i \psi}}.
\eeq
To re-write the $6d$ metric in terms of these coordinates we should first define 
one-forms\footnote{Notice that $Z_2$ and $Z_3$ are defined only up to a multiplicative constant. The 
forms $\eta_i$ and $\widetilde{\eta}_i$, however, are independent of these constants.}
$\eta_i=d (\ln Z_i)$
and $\widetilde{\eta}_i$:
\beq
\label{eq:etaeta}
\widetilde{\eta}_1 = \eta_1, \qquad
\widetilde{\eta}_2 = \eta_2 - \cos \theta \eta_1, \qquad
\widetilde{\eta}_3 = \eta_3 - y \eta_2 - \cos \theta (1-y) \eta_1.
\eeq
This enables us to recast the metric in the following neat form:
\beq
\label{eq:metric-eta}
d s^2_{(6)} = d r^2 + r^2 \left( \frac{1-y}{6} \widetilde{\eta}_1 \overline{\widetilde{\eta}_1}
                                   + \frac{w(y)q(y)}{36} \widetilde{\eta}_2 \overline{\widetilde{\eta}_2}
                                   + \frac{1}{9} \widetilde{\eta}_3 \overline{\widetilde{\eta}_3} \right).
\eeq
The variables $Z_i$ are singular. For instance, $Z_1$ diverges when $\theta=\pi$. The regular combinations
of $Z_1$, $Z_2$ and $Z_3$ give rise to the aforementioned variables $a_j$, $b_j$ and $c_j$ via the relations:
\beq
a_j = Z_1^{\frac{1}{2}(p-q)-j} Z_2^{-\frac{1}{6 \ell}} Z_3^{\frac{1}{6 y_2\ell}} , \quad
b_0 = Z_1^{-1} Z_3, \quad   b_1 = Z_3, \quad        b_2 = Z_1 Z_3, \quad
c_j = Z_1^{\frac{1}{2}(p+q)-j} Z_2^{\frac{1}{6 \ell}} Z_3^{-\frac{1}{6 y_1\ell}}.
\eeq
As a consistency check one can easily verify that these variables satisfy exactly the same relations
as the variables introduced in (\ref{eq:abc}). In doing so various relations between $y_1$, $y_2$
and $\ell$ might be useful. These relations are collected in Appendix \ref{AppC}. Using these relations it is also possible to
express $Z_1$, $Z_2$ and $Z_3$ in terms of $\z_i$'s. We find that:
\beq
Z_1 = \frac{\z_2}{\z_1}, \qquad
Z_2 = \z_1 \z_2 \z_3^\frac{1}{y_1} \z_4^\frac{1}{y_2}, \qquad
Z_3 = \z_1 \z_2 \z_3 \z_4.
\eeq
We found the connection between the singular holomorphic coordinates $(Z_1,Z_2,Z_3)$
and the K\"ahler quotient coordinates $(\z_1,\z_2,\z_3,\z_4)$. While the former uniquely determine
(through the definition of the $\eta_i$'s and (\ref{eq:etaeta}))
the $1$-forms $\widetilde{\eta}_i$  in the metric (\ref{eq:metric-eta}), the latter are related to 
$u_1$, $u_2$, $v_1$ and $v_2$ (through the definitions (\ref{eq:uuvvzzzz})). Since for $q=1,2$ we have found explicit
maps from the space parameterized by $u_{1,2}$ and $v_{1,2}$ to $S^3 \times S^2$, there is also a direct way to
re-write the forms $\widetilde{\eta}_1$, $\widetilde{\eta}_2$ and $\widetilde{\eta}_3$ in terms of the  $S^3 \times S^2$
coordinates arriving eventually at a $5d$ $Y^{p,q}$ metric in an explicit $S^3 \times S^2$ form.
The final result, however, is extremely long and complicated and we will not report it here.
The main reason for this is the proper normalization of $u_1$, $u_2$, $v_1$ and $v_2$, which we have not addressed yet.

\begin{figure}
\begin{center}
\includegraphics[width=0.7\textwidth]{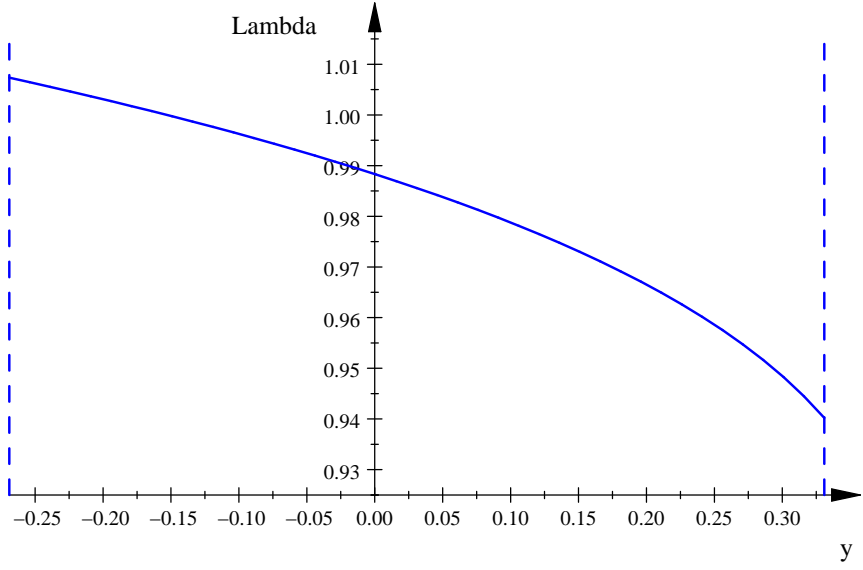}
\caption{The plot shows a numerical solution of (\ref{eq:Ly}) for $p=5$, $q=2$. We see that $\Lambda(y)$ is
 a monotonically decreasing function of $y$ and so for a fixed $y$ there is a unique solution of (\ref{eq:Ly}). 
 One can also directly check that the boundary points
 $\Lambda(y_1) \approx 1.0073534$ and $\Lambda(y_2) \approx 0.93921856$ on the graph are indeed solutions of 
 the equation (\ref{eq:Ly}). }
\label{Pict1}
\end{center}
\end{figure}

As we have already mentioned, the expressions for $\z_i$'s in terms of the metric coordinates $r$, $y$, $\theta$, $\phi$,
$\psi$ and $\alpha$ will inevitably include a complex parameter. Its absolute value should be fixed 
by the $D$-term condition and the phase corresponds to the $U(1)_K$ gauge.
Let us denote the absolute value by $\Lambda$ and the gauge parameter by $\lambda$. Then the expressions read:
\begin{eqnarray}
\label{eq:z1z2z3z4}
\z_1 &=& \Lambda^p \cdot \cos \frac{\theta}{2} e^{-\frac{i}{2} \phi + i p \lambda}, \nonumber \\
\z_2 &=& \Lambda^p \cdot \sin \frac{\theta}{2} e^{\frac{i}{2} \phi + i p \lambda}, \nonumber \\
\z_3 &=& \Lambda^{-p+q} \cdot r^{- \frac{3 y_1}{y_2-y_1} }  (y-y_1)^\half (y_3-y)^{\frac{1}{2} \frac{y_3-1}{1-y_1}}
            e^{\frac{i}{p} \left( \frac{p-q}{2}  \psi - \frac{\alpha}{\ell} \right) - i (p-q)\lambda}, \nonumber \\
\z_4 &=& \Lambda^{-p-q} \cdot r^{ \frac{3 y_2}{y_2-y_1} }  (y_2-y)^\half (y_3-y)^{\frac{1}{2} \frac{y_3-1}{1-y_2}}
            e^{\frac{i}{p} \left( \frac{p+q}{2}  \psi + \frac{\alpha}{\ell} \right) - i (p+q)\lambda}.
\end{eqnarray}
Notice that if we define $u_1$, $u_2$, $v_1$ and $v_2$ as in (\ref{eq:uuvvzzzz}) 
with a unit normalization factor,
and then normalize
the coordinates as in (\ref{eq:length-one}), we find that $\Lambda=1$ (from the $u$-normalization) and
the radial coordinate $r$ is a complicated function of $y$ (from the $v$-normalization). Although this approach is certainly 
legitimate, it does not correspond to the $Y^{p,q}$ $5d$ metric (\ref{eq:5d}), which is defined as an $r=1$ ``slice"
of the cone. Thus, in order to stick to the $r=1$ choice, we have to set $r=1$ and to modify the original definition: (\ref{eq:uuvvzzzz}) to
\beq
u_1 \equiv \Lambda^{-p} \z_1, \qquad 
u_2 \equiv \Lambda^{-p} \z_2, \qquad 
v_{1} \equiv \Lambda^{-p} \sqrt{1- \frac{q}{p}} \,\, \overline{\z}_{3}, \qquad
v_{2} \equiv \Lambda^{-p} \sqrt{1+ \frac{q}{p}} \,\, \overline{\z}_{4},
\eeq
which looks exactly like (\ref{eq:uuvvzzzz}) except for the over-all normalization factor $\Lambda^{-p}$.
Now the $D$-term condition (\ref{eq:Dterm}) and the unit-length normalization (\ref{eq:length-one}) lead to the same
equation for $\Lambda$ in terms of $y$:
\beq
\label{eq:Ly}
\Lambda(y)^{4p} = \left(1- \frac{q}{p}\right) (y-y_1) (y_3-y)^{\frac{y_3-1}{1-y_1}} \cdot \Lambda(y)^{2q} +
 	             \left(1+ \frac{q}{p}\right) (y_2-y) (y_3-y)^{\frac{y_3-1}{1-y_2}} \cdot \Lambda(y)^{-2q}. 
\eeq
To cast the metric in the explicit $S^3 \times S^2$ form we will need an analytic solution of the above equation,
which we believe does not exist. We have, however, analyzed the equation numerically for various $p$'s and $q$'s
verifying that it possesses only a single 
solution for $\Lambda=\Lambda(y)$ in the $y_1\leqslant y \leqslant y_2$ range, making the whole normalization procedure well-defined.
For $(p,q)=(5,2)$ the function $\Lambda(y)$ is presented in Figure \ref{Pict1}.

To conclude, we have established a relation between the metric coordinates (\ref{eq:5d}) and the $S^3 \times S^2$
coordinates of the previous section. The explicit result, unfortunately is very complicated due to the normalization issue.

Let us end this section with a remark regarding the regularity of $a_j$'s, $b_i$'s and $c_j$'s.
The phases of these variables are as follows:
\begin{eqnarray}
\label{eq:phases}
&& \textrm{Arg} (a_j) = \frac{1}{2} (p-q-2j) \phi + \frac{1}{2} (p-q) \psi -\frac{\alpha}{\ell}
                                                                            =-\tau+p (\psi + \phi) - j\phi,  \nonumber \\
&& \textrm{Arg} (b_0,b_1,b_2) = \psi-\phi,\psi,\psi+\phi,                                       \nonumber \\
&& \textrm{Arg} (c_j) = \frac{1}{2} (p+q-2j) \phi + \frac{1}{2} (p+q) \psi +\frac{\alpha}{\ell}
                                                                            =\tau-j \phi. 
\end{eqnarray}
Notice that only if $\phi$, $\psi$ and the angle $\tau$ introduced in (\ref{eq:alphaNEW}) are all $2\pi$-periodic, 
the phases in (\ref{eq:phases})
are well defined. Clearly, this check is equivalent to the calculation explained around (\ref{eq:alphaNEW}).
For instance, for $\theta=0$ only $a_0$, $b_0$ and $c_0$ are non-zero and their algebra ($a_0c_0=b_0^p$) 
properly describes a cone over the lens space $L(p;p-1)$. Analogously, for $y=y_1$ only $c_j$'s do not vanish and their algebra 
corresponds to the $L(p+q;1)$ cone, while for $y=y_2$ the 
variables $a_j$'s reproduce the $L(p-q;1)$  algebra.

\section{Normalizing the RR flux} \label{normsec}

Now we are in a position to use the results of the previous two sections to calculate the flux through the three-sphere
for $q=1$ and $q=2$. The RR $3$-form $F_3$ is the real part of the self-dual $(2,1)$ form $G_3$ found 
in \cite{HEK}. The RR $2$-form potential is given by:
\beq
\label{eq:A2}
A_2 = \frac{K}{3} \left( \frac{1}{1-y}\d \alpha \wedge \d \psi + \frac{1}{6} \cos \theta \, \d \psi \wedge \d \phi
 - \frac{y  \cos \theta}{1-y} \d \alpha \wedge \d \phi \right)
 \quad \textrm{with} \quad
K = \frac{9}{8 \pi^2} (p^2-q^2). 
\eeq
In what follows we will compute 
the integral $\int_{S^3} F_3$ verifying that it yields $1$ for the above choice of the constant $K$.  The homology class of the calibrated lens space $L(j;1)$ is equal to $j$ divided by this integral, following an argument in \cite{EKK}.

In this section we will find it convenient to parametrize the coordinates $u$ and $v$ as:
\beq
\label{eq:xi}
(u_1,u_2) = \left( \cos \frac{\theta}{2} \,\, e^{i \phi_1}, \,\,\, \sin\frac{\theta}{2} \,\, e^{i \phi_2} \right) 
\qquad \textrm{and} \qquad
(v_1,v_2) = \left( \cos \frac{\xi}{2} \,\, e^{i \phi_3}, \,\,\, \sin\frac{\xi}{2} \,\, e^{i \phi_4} \right).
\eeq
Here $\xi$ is a well-defined, though complicated, function of $y$. To write $\xi(y)$ explicitly one would need an
analytic solution of (\ref{eq:Ly}), which, we guess, does not exist. In what follows, however, 
it will be enough to know only the range of $\xi$.  A simple substitution shows that $\xi=0$ for $y=y_2$
and  $\xi=\pi$ for $y=y_1$. The angles $\phi_i$ are, of course, gauge dependent.
A basis of gauge invariant combinations is:
\beq
\label{eq:phi1234}
\phi= \phi_2-\phi_1, \qquad
\psi= \phi_1+\phi_2-\phi_3-\phi_4, \quad \textrm{and} \quad
\tau= (p+q) \phi_2 - p \phi_4,
\eeq
where we have used (\ref{eq:uuvvzzzz}), (\ref{eq:alphaNEW}) and (\ref{eq:z1z2z3z4}).
Again, we see that $\tau$ is $2 \pi$-periodic as was asserted in the previous section.

\subsection{$q=1$}

\begin{figure}[t]
\label{Pict2}
\setlength{\unitlength}{1.7pt}
\qquad \quad
\centering
\begin{picture}(140,140)
\put(120,60){\vector(-1,-1){10}}
\put(55,117){\vector(1,-1){10}}
\put(-5,50){\vector(1,-1){10}}
\put(45,25){\vector(1,-1){10}}
\put(40,122){$L(p-q;1)$}
\put(30,30){$L(p+q;1)$}
\put(-30,55){$L(p;p-1)$}
\put(115,65){$L(p;p-1)$}
\put(73,73){\Large $S^3$}
\put(10,10){\vector(0,1){110}}
\put(10,10){\vector(1,0){110}}
\put(10,0){$0$}
\put(-35,10){$y=y_1$, $\xi=\pi$}
\put(130,8){$\theta$}
\put(8,128){$y$}
\put(105,0){$\pi$}
\put(-35,105){$y=y_2$, $\xi=0$}
\put(10,10){\line(1,0){95}}
\put(10,10){\line(0,1){95}}
\put(105,10){\line(0,1){95}}
\put(10,105){\line(1,0){95}}
\linethickness{0.7mm}
\qbezier(10,105)(80,80)(105,10)
\end{picture}
\caption{The $S^3$ embedding given by $s_2=0$ (or equivalently $\xi(y)=\theta$ and $\psi=0$ ). 
The solid curve is the three-sphere. The $2 \pi$-periodic angles along the $3$-sphere are $\phi$ and $\tau$.}
\end{figure}
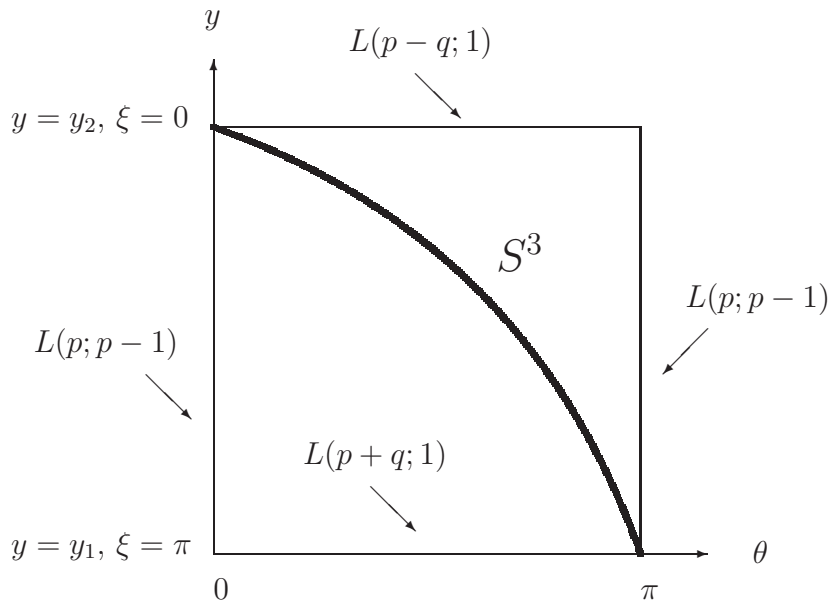

To calculate the flux through the three-sphere we first have to fix
the two-sphere coordinate. We will choose $s_2=0$. With this choice 
the second equation in (\ref{eq:XSq1}) implies that:
\beq
u_2 \overline{v}_1 = \overline{u}_1 v_2.
\eeq
With the help of the equations (\ref{eq:xi}) and (\ref{eq:phi1234}) we find that the $S^3$ 
embedding is given by:
\beq
\xi = \theta
\qquad  \textrm{and} \qquad
\psi=0.
\eeq
Here the first equation yields $y$ as a function of $\theta$. The schematic form
of the embedding in the $(y,\theta)$-plane is depicted in Figure \ref{Pict2}.
Using the first equation in (\ref{eq:XSq1}) we can also find the $S^3$ coordinates $x_1$ and $x_2$
in terms of the metric coordinates $\theta$, $\phi$ and $\tau$:
\beq
(x_1,x_2) = \left( \cos \frac{\theta}{2} \,\, e^{i (\tau-\phi)}, \,\,\, \sin\frac{\theta}{2} \,\, e^{i \tau} \right) . 
\eeq
This provides an additional check that $\tau$ is $2 \pi$-periodic.

We are now ready to integrate the flux $F_3$ over $S^3$.  As $F_3$ is closed, it may be written almost everywhere as $dA_2$ where $A_2$ is given (\ref{eq:A2}).  $F_3$ is not exact, so $A_2$ is necessarily singular.  This is similar to the integral of the magnetic field over a surface linking a monopole.  The magnetic field strength may be written as the exterior derivative of a vector potential which diverges at certain gauge-dependent points called Dirac strings.  As the field strength is exact away from the Dirac string, after an application of Stokes' theorem this region does not contribute to the integral.  In fact, the entire integral of the magnetic field comes from the Dirac string itself.  More precisely, one can consider a small loop around the string, integrate the potential around the loop and then take the limit in which the loop shrinks away. Although the gauge potential grows as the loop shrinks, the integral converges. By Stokes' theorem this limit of the integral of the vector potential around the Dirac string is equal to the integral of the magnetic field on the entire surface. In the present case we will not need to take a limit, because our Dirac strings themselves will already be 2-dimensional.  We will thus refer to them as Dirac surfaces.

Here we have two Dirac surfaces, at the endpoints $y=y_1$ and $y=y_2$. Stokes' theorem tells us that the integral of $F_3$ over the three-sphere is just the sum of the integrals of $A_2$ over the two Dirac surfaces. Remarkably, since $\psi=0$, only the last term
in (\ref{eq:A2}) contributes to the integral.  Using (\ref{eq:alphaNEW}) we obtain:
\beq
\int_{S^3} F_3 = - \frac{K}{3} \int_{\xi=\theta, \psi=0}  \left.  \frac{y  \cos \theta}{1-y} 
         \d \alpha \wedge \d \phi\ \right\vert^{\theta=0}_{\theta=\pi} =
- \frac{4 \pi^2 \ell}{3} K \left( \frac{y_2}{1-y_2} + \frac{y_1}{1-y_1} \right) = q =1
\eeq
as expected.	
Here we used (\ref{eq:alphaNEW}), the explicit form of $K$ 
in (\ref{eq:A2}) as well as the first two relations involving $y_1$ and $y_2$ collected in Appendix \ref{AppC}.

\subsection{$q=2$}

Here we will again  consider the $s_2=0$ embedding of the three-sphere.
From (\ref{eq:SUW}) we obtain:
\beq
\label{eq:q2uw}
u_2 \overline{w}_1^k = \overline{u}_1 \frac{w_2^{k+1}}{|w_2|},
\qquad \textrm{where} \quad 
w_1 = u_1 \overline{v}_1 + \overline{u}_2 v_2 \quad \textrm{and} \quad w_2 = u_2 \overline{v}_1 - \overline{u}_1 v_2.
\eeq
In contrast with the $q=1$ case, now there is no simple relation between $\theta$ and $\xi(y)$.
However, for the flux calculation no such relation is needed. Indeed, as we have argued, since $F_3=\d A_2$ is closed, the only 
non-zero contributions to the integral $\int_{S^3} F_3$ come from the surfaces where the form 
$A_2$ is ill-defined. These surfaces are given by $\theta=0,\pi$ or $y=y_{1,2}$, 
where various cycles corresponding to the
angles $\phi$, $\psi$ and $\tau$ collapse.

\begin{figure}[t]
\label{Pict3}
\setlength{\unitlength}{1.7pt}
\qquad
\centering
\begin{picture}(140,140)
\put(10,105){\circle{15}}
\put(105,105){\circle{15}}
\put(10,10){\vector(0,1){110}}
\put(10,10){\vector(1,0){110}}
\put(10,0){$0$}
\put(-45,10){$y=y_1$, $\xi=\pi$}
\put(130,8){$\theta$}
\put(8,128){$y$}
\put(105,0){$\pi$}
\put(55,0){$\pi/2$}
\put(-45,105){$y=y_2$, $\xi=0$}
\put(10,10){\line(1,0){95}}
\put(10,10){\line(0,1){95}}
\put(105,10){\line(0,1){95}}
\put(10,105){\line(1,0){95}}
\linethickness{0.7mm}
\qbezier(10,105)(50,57.5)(57.5,10)
\qbezier(105,105)(65,57.5)(57.5,10)
\qbezier(10,105)(30,90)(57.5,105)
\qbezier(57.5,105)(85,90)(105,105)
\end{picture}
\caption{The $S^3$ embedding defined by $s_2=0$ for $q=2$. The area surrounded by the solid curve 
corresponds to the three-sphere. There are four points
in the intersection of the three-sphere and the rectangle defined by $\theta=0,\pi$ and $y=y_{1,2}$. 
Each point corresponds to a Dirac surface that contributes to the flux. 
At the circled points we need to slightly deform the Dirac surfaces.}
\end{figure}
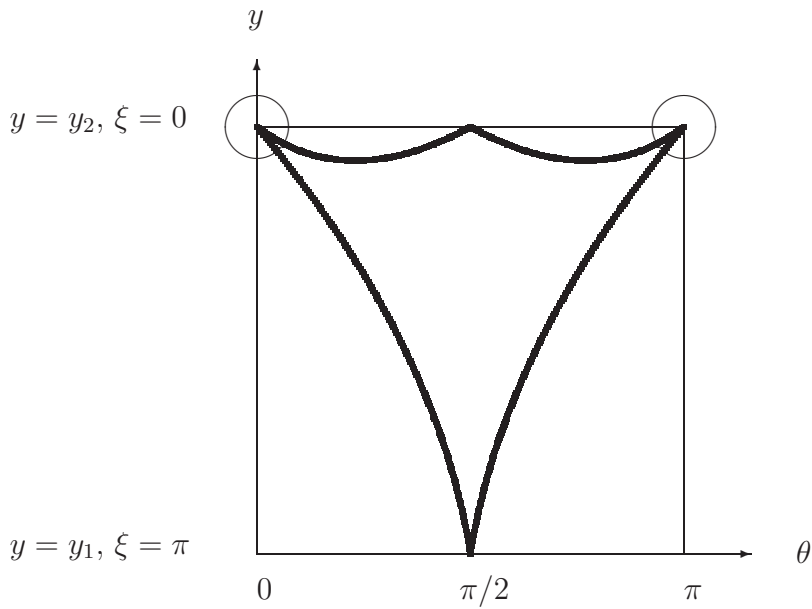

Thus we need to find all possible intersections of (\ref{eq:q2uw}) with the $3d$ surfaces $u_{1,2}=0$
and  $v_{1,2}=0$.
It appears that the results are slightly different for $k>1$ and $k=1$. We will relegate the $k=1$
case to Appendix \ref{AppD}, assuming that $k > 1$ in the rest of the section.
Substituting (\ref{eq:xi}) into (\ref{eq:q2uw}) we find four Dirac surfaces:
\beq
\label{eq:sol}
1. \,\,(\theta=\frac{\pi}{2}, \xi=\pi) \qquad
2. \,\,(\theta=\frac{\pi}{2}, \xi=0) \qquad
3. \,\,(\theta=0, \xi=0) \qquad
4. \,\,(\theta=\pi, \xi=0).
\eeq
As we will see, for each one of the solutions  only one periodic coordinate is constrained.
Thus (\ref{eq:sol})
describes four different $2d$ tori inside $Y^{p,2}$. These tori are the Dirac surfaces 
and so the integral of $F_3$ over the 3-sphere will be the sum of the integrals of $A_2$ over the surfaces.  The situation is then slightly more complicated than it was in the $q=1$ case, where we had only two
surfaces at $(\theta=0, y=y_2) $ and $(\theta=\pi, y=y_1) $. 
A typical form of the $s_2=0$ embedding for $q=2$ is depicted on Figure \ref{Pict3}.

The first solution in (\ref{eq:sol}) corresponds to $|u_1|=|u_2|$ and $v_1=0$.  Therefore the $w$'s are just:
\beq
w_1=\overline{u}_2v_2,\qquad w_2=-\overline{u}_1v_2.
\eeq
Substituting this into (\ref{eq:q2uw}) we find that:
\beq
(k+2) \phi_1 + (k+1) \phi_2 - (2k+1) \phi_4 = (k+1 + 2M) \pi 
\qquad \textrm{for} \quad M \in \Z,
\eeq
which with the help of (\ref{eq:phi1234}) implies that:
\beq
\tau - (k+2) \phi =  (k+1 + 2M) \pi.
\eeq
Similarly, for the second point in (\ref{eq:sol}):
\beq
|u_1|=|u_2|, \qquad v_2=0,\qquad w_1=u_1\overline{v}_1,\qquad w_2=u_2\overline{v}_1
\eeq
and so  we have:
\beq
(k-1) \phi_1 + k \phi_2 - (2k+1) \phi_3 = 2\pi M ,
\eeq
which by (\ref{eq:phi1234}) implies that:
\beq
  (k+2) \phi + (2k+1)\psi  -\tau= 2 \pi M .
\eeq
Next let us consider the $(\theta=0, \xi=0)$ point.
It appears that in order to find a corresponding angular coordinate constraint
we have to slightly deform the surface. This happens because the equation (\ref{eq:q2uw})
is singular at the Dirac surface. 
Clearly, such a deformation cannot change the final result for the flux.
We found that the following Ans\"atz does the job:
\beq
u_1 = e^{i \phi_1}, \qquad
u_2 = \epsilon^k e^{i \phi_2}, \qquad
v_1 = e^{i \phi_3}, \qquad
v_2 = \epsilon e^{i \phi_4},
\eeq
where $\epsilon$ is an infinitesimal deformation parameter.
Substituting this into (\ref{eq:q2uw}) and keeping only terms of order $\epsilon^k$, we arrive at:
\beq
2\phi_1 + \phi_2 + k \phi_3 - (k+1) \phi_4 =  (k+1 + 2M)\pi,
\eeq
so 
\beq
 (k+2) (\phi+\psi) -\tau =  (k+1 + 2M) \pi.
\eeq
Finally, for $(\theta=\pi, \xi=0)$ the deformation is:
\beq
u_1 = \epsilon^k e^{i \phi_1}, \qquad
u_2 = e^{i \phi_2}, \qquad
v_1 = e^{i \phi_3}, \qquad
v_2 = \epsilon e^{i \phi_4},
\eeq
which leads to:
\beq
\phi_1 + (k+1) \phi_3 - k \phi_4 = 2 \pi M,
\eeq
and
\beq
  (k+2)\phi + (k+1) \psi  -\tau =  2 \pi M.
\eeq
Remarkably, for each one of the solutions in (\ref{eq:sol})  the $\tau$ angle can 
be expressed uniquely in terms of $\psi$ and $\phi$. This means that for all the
Dirac surfaces in (\ref{eq:sol})  $\psi$ and $\phi$ are well-defined $2\pi$ periodic coordinates.
It is now a straightforward exercise to compute the four contributions to the integral.
These are:
\beq
\frac{2 \pi^2 \ell K}{3} \frac{1}{1-y_1}, \quad
\frac{2 \pi^2 \ell K}{3} \frac{1}{1-y_2}, \quad
\frac{2 \pi^2 \ell K}{3} \left( -\frac{1-3y_2}{1-y_2} +\frac{1}{3 \ell}  \right), \quad
\frac{2 \pi^2 \ell K}{3} \left( -\frac{1+y_2}{1-y_2} -\frac{1}{3 \ell} \right) 
\eeq
respectively. An important question we have to address is the orientation
of the contributions.  The easiest way to fix the relative orientation is to consider ``probe"
forms $\d \psi \wedge \d \phi$, $\d \tau \wedge \d \phi$ and $\d \tau \wedge \d \psi$. 
Although these forms are closed and as such 
cannot contribute to the integral $\int_{S^3} F_3 $, all of them still have non-vanishing contributions near
the Dirac surfaces, that must eventually sum to zero for each form separately. Using this requirement we can fix 
relative orientations of
all possible terms in $A_2$.

We are finally ready to calculate the flux:
\beq
\int_{S^3} F_3  = \frac{2 \pi^2 \ell K}{3} \left( \frac{1}{1-y_1}-\frac{1}{1-y_2} \right)
 + \frac{4 \pi^2 \ell K}{3} \frac{y_2}{1-y_2}
 = -\frac{p}{2} + \frac{p+q}{2} = \frac{q}{2}=1.
\eeq

\section{Conclusions}

In this paper we have constructed an explicit homeomorphism between the $Y^{p,q}$ spaces
and the product space $S^3 \times S^2$ for $q=1$ and $2$. There are plenty of open questions that deserve 
further investigation.  An immediate direction, of course, is to find trivializations for higher $q$'s
as well as for the $L^{a,b,c}$ spaces. We pointed out in the paper that the main obstacle to extending our approach to the 
cases $q>2$ is the fact that $\Z_q$ will no longer be a normal subgroup of $SU(2)$.

Before extending the trivialization to yet more infinite families, one may wish to exploit the trivializations that we have already found.  One obvious result is that one can use the trivialization to identify $S^3$ representatives of the third homology generator.  A large family of representatives is given, for example, by choosing an element of $S^2$ for each element of $S^3$.  One can then wrap branes around these cycles, corresponding to baryonic operators in the dual gauge theory. The baryonic charge is given by the homology class, and so these will be operators of charge one.  One can then use the metric, at least numerically, to calculate the volumes of these branes which will determine the R-charges of these operators.
Remarkably, these baryon operators should be constructed from both chiral and anti-chiral superfields
since the cone over the three-sphere is not holomorphic and therefore non-supersymmetric.

One may also use this trivialization to construct orbifolds of $Y^{p,q}$, as was done for the conifold in \cite{EK}.  The identification of the $S^3$ also allows one to geometrically construct a deformation of the tip of the cone over $Y^{p,q}$ in which the singularity is replaced with an $S^3$ homotopic to that of $Y^{p,q}$. 
The deformed $6d$ space can then be used to construct a $10d$ supergravity background by analogy with the 
Klebanov-Strassler solution \cite{KS} based on the deformed conifold $6d$ geometry of \cite{Candelas:1989js}.
While such a deformation cannot be supersymmetric \cite{Berenstein:2005xa,Franco:2005zu,Bertolini:2005di}, 
it may nonetheless be interesting to see what it corresponds to in the field theory. 
The proposed solution will describe a flow from the superconformal theory dual to the $AdS_5 \times Y^{p,q}$
geometry to some non-supersymmetric gauge theory.

\section*{Acknowledgements}

 We would like to thank Ofer Aharony, Riccardo Argurio,
Andrea Brini, Cyril Closset, 
Oleg Khasanov, 
Dario Martelli, Yaron Oz, Cobi Sonnenschein, 
James Sparks, Shimon Yankielowicz
and especially Daniel Persson
for fruitful discussions.

\appendix





\section{Homology from the Gysin sequence} 
\label{Homol}

Given the Chern class of a circle bundle and the cohomology of the base $M$ one can determine
the cohomology of the total space $E$, or given partial information about all three one can often determine the rest.  
This is not surprising, as the cohomology of $E$ is completely characterized by that of $M$.  
The relation between the cohomology groups however is quite simple, 
they are related by a long exact sequence known as the Gysin sequence.  

The Gysin sequence is
\beq
...\stackrel{\pi^*}{\longrightarrow} \H^n(E)\stackrel{\pi_*}{\longrightarrow}\H^{n-1}(M)\stackrel{c\cup}{\longrightarrow}\H^{n+1}(M)\stackrel{\pi^*}{\longrightarrow}\H^{n+1}(E)\stackrel{\pi_*}{\longrightarrow} ...
\eeq
where $\pi^*$ and $\pi_*$ are the pullback and pushforward of the projection map $\pi:E\longrightarrow M$ 
and $c\cup$ is the cup product with the Chern class.  This long exact sequence is exact, 
meaning that the image of each map is the kernel of the next.  Using this fact we can compute 
the homology of $Y^{p,q}$.

$Y^{p,q}$ is a circle bundle over $M=S^2\times S^2$.  The Chern class is an element of
\beq
\H^2(S^2\times S^2)=\Z^2
\eeq
and so is a pair of integers, $p$ and $q$.  We can find the cohomology groups of a general $Y^{p,q}$ 
using the Gysin sequence, even when $p$ and $q$ are not relatively prime.

The first non-trivial part of the Gysin sequence is
\beq
0\stackrel{(p,q)\cup}{\longrightarrow}\H^0(S^2\times S^2)=\Z\stackrel{\pi^*}{\longrightarrow} \H^0(Y^{p,q})\stackrel{\pi_*}{\longrightarrow}\H^{-1}(S^2\times S^2)=0
\eeq
and so the pullback $\pi^*$ is an isomorphism, yielding
\beq
\H^0(Y^{p,q})=\H^0(S^2\times S^2)=\Z
\eeq
which means that $Y^{p,q}$ is connected.

The next piece is
\bea
0&\stackrel{(p,q)\cup}{\longrightarrow}&\H^1(S^2\times S^2)=0\stackrel{\pi^*}{\longrightarrow} \H^1(Y^{p,q})\stackrel{\pi_*}{\longrightarrow}\H^{0}(S^2\times S^2)=\Z\nonumber\\
&\stackrel{(p,q)\cup}{\longrightarrow}&\H^2(S^2\times S^2)=\Z^2\stackrel{\pi^*}{\longrightarrow} \H^2(Y^{p,q})\stackrel{\pi_*}{\longrightarrow}\H^{1}(S^2\times S^2)=0.
\eea
Again, assuming that $p$ and $q$ are not both equal to zero, we find that $(p,q)\cup$ has no kernel and so 
\beq
\H^1(Y^{p,q})=0.
\eeq
However the image of $(p,q)\cup$ in $\H^2(S^2\times S^2)=\Z^2$ is more complicated.  
Again it is only a proper sublattice of $\Z^2$, but this time it misses an entire free group $\Z$ 
plus anything which when multiplied by a constant gives the element $(p,q)$.  
Such elements form a finite cyclic subgroup whose order is $\gcd(p,q)$, the greatest 
common divisor of $p$ and $q$.
As $p$ and $q$ are relatively prime, $\gcd(p,q)=1$, and so this cyclic group is trivial.
Therefore
\beq
\H^2(Y^{p,q})=\Z.
\eeq

The next useful piece is
\bea
0=\H^3(S^2\times S^2)&\stackrel{\pi^*}{\longrightarrow}& \H^3(Y^{p,q})\stackrel{\pi_*}{\longrightarrow}\H^{2}(S^2\times S^2)=\Z^2
\stackrel{(p,q)\cup}{\longrightarrow}\H^4(S^2\times S^2)=\Z\nonumber\\&\stackrel{\pi^*}{\longrightarrow}& \H^4(Y^{p,q})\stackrel{\pi_*}{\longrightarrow}\H^{3}(S^2\times S^2)=0.
\eea
The kernel of
\beq
(p,q)\cup:\H^2(S^2\times S^2)=\Z^2\longrightarrow \H^4(S^2\times S^2)=\Z \label{ja:ymap}
\eeq
is $\Z$, which is generated by $(q,-p)/\gcd(q,p)$ and so
\beq
\H^3(Y^{p,q})=\Z.
\eeq
The image of (\ref{ja:ymap}) on the other hand is not all of $\Z$, but just the subset consisting of 
numbers with are sums of multiples of $p$ by multiples of $q$, which is the same as the subset 
of multiples of $\gcd(p,q)=1$.  This subset is the kernel of the next map, the pullback to $\H^4(Y^{p,q})$, 
and so the image of that map is 
\beq
\H^4(Y^{p,q})=\Z_{\gcd(p,q)}=0.
\eeq

The last useful part of the Gysin sequence is
\beq
0=\H^5(S^2\times S^2)\stackrel{\pi^*}{\longrightarrow} \H^5(Y^{p,q})\stackrel{\pi_*}
{\longrightarrow}\H^{4}(S^2\times S^2)=\Z
\stackrel{(p,q)\cup}{\longrightarrow}\H^6(S^2\times S^2)=0
\eeq
and so
\beq
\H^5(Y^{p,q})=\H^4(S^2\times S^2)=\Z
\eeq
establishing that the $Y^{p,q}$ spaces are orientable.

Now that we have the cohomology of the spaces $Y^{p,q}$, and we know that they are compact 
and orientable, we may get the homology from Poincar\'e duality
\beq
\H_0(Y^{p,q})=\H_2(Y^{p,q})=\H_5(Y^{p,q})=\Z   \hsp \H_1(Y^{p,q})=0 \hsp  \H_3(Y^{p,q})=\Z\ \hsp \H_4(Y^{p,q})=0.
\eeq
Substituting the homology of the $3$-sphere
\beq
\H_0(S^3)=\H_3(S^3)=\Z \hsp \H_1(S^3)=0 \hsp \H_2(S^3)=0
\eeq
and the 2-sphere
\beq
\H_0(S^2)=\H_2(S^2)=\Z \hsp H_1(S^2)=0
\eeq
into the K\"unneth formula
\beq
\H_p(S^2\times S^3)=\oplus_i \H_i(S^2)\otimes \H_{p-i}(S^3),
\eeq
one finds that $S^2 \times S^3$ has the same homology groups as $Y^{p,q}$.  
Note that the K\"unneth formula has no Tor corrections because the sphere has no torsion homology.

\section{The homotopy groups}
\label{Homot}

The fundamental group of $Y^{p,q}$ can be calculated again using the fact that it is a circle 
bundle over $S^2\times S^2$, and using the long exact sequence of homotopy groups of a fibration.  
We will calculate it and show that for co-prime $p$ and $q$ 
it is equal to that of $S^2\times S^3$.  The long exact sequence for homotopy groups  of 
a fibration $S^1\longrightarrow Y^{p,q}\longrightarrow S^2\times S^2$ is (see \emph{Switzer} 4.7 for example)
\beq
...\longrightarrow\pi_{n+1}(S^2\times S^2)\stackrel{\partial}{\longrightarrow}\pi_n(S^1)  \stackrel{i_*}{\longrightarrow}\pi_n(Y^{p,q})\stackrel{p_*}{\longrightarrow}\pi_n(S^2\times S^2)\longrightarrow...
\eeq
where $p$ is the projection map $p:Y^{p,q}\longrightarrow S^2\times S^2$, $i$ is the inclusion 
of the fiber into the total space and $\partial$ roughly measures the transition function.

Using the fact that
\beq
\pi_2(S^1)=\pi_1(S^2\times S^2)=0
\eeq
we may isolate the part of this sequence between the two vanishing terms
\beq
0\longrightarrow\pi_2(Y^{p,q})\stackrel{p_*}{\longrightarrow}\pi_2(S^2\times S^2)=\Z^2\stackrel{\partial}{\longrightarrow}\pi_1(S^1)=\Z\stackrel{i_*}
{\longrightarrow}\pi_1(Y^{p,q})\longrightarrow 0. \label{hseq}
\eeq
The two unknown terms are now completely determined by the fact that the boundary map is just 
given by the Chern class
\beq
\label{bordo}
\partial(a,b)=pa+qb. 
\eeq
Since $p$ and $q$ are co-prime, the image of the boundary map therefore consists of all integers.  
This group is therefore the kernel of the map $i_*:\pi_1(S^1)\longrightarrow\pi_1(Y^{p,q})$ and so
\beq
\pi_1(Y^{p,q})=\Z_{\gcd(p,q)}=0
\eeq
as advertised.  

We may also use (\ref{hseq}) to obtain the second homotopy group of $Y^{p,q}$.  The kernel of the
 boundary map (\ref{bordo}) is a subgroup of $\Z^2$.  It contains all elements of the form $(kq,-kp)$ 
so it is at least one-dimensional, but if either $p$ or $q$ is non-zero then it does not contain all 
elements, so it is at most one-dimensional.  Therefore the kernel of $\partial$ is $\Z$, which is 
the image of the injective map $p_*:\pi_2(Y^{p,q})\longrightarrow \pi_2(S^2\times S^2)$.  Therefore
\beq
\pi_2(Y^{p,q})=\Z.
\eeq
This agrees with the second homotopy group of the product of the 2-sphere and 3-sphere, as the second 
homotopy group of the 3-sphere is the trivial group and that of the 2-sphere is $\Z$.

All of the higher homotopy groups $\pi_k(Y^{p,q})$ are easily found because the corresponding homotopy 
groups of the circle are trivial
\beq
0=\pi_k(S^1)\longrightarrow\pi_k(Y^{p,q})\stackrel{p_*}{\longrightarrow}\pi_k(S^2\times S^2)\stackrel{\partial}{\longrightarrow}\pi_{k-1}(S^1)=0.
\eeq
The exactness of this sequence implies that the higher homotopy groups of $Y^{p,q}$ are isomorphic 
to those of $S^2\times S^2$
\beq
\pi_k(Y^{p,q})=\pi_k(S^2\times S^2)\hsp k>2.
\eeq
This can be further simplified using the K\"unneth formula
\beq
\pi_k(M\times N)=\pi_k(M)\oplus\pi_k(N)
\eeq
so that the homotopy groups of $Y^{p,q}$ can be expressed in terms of those of the 2-sphere
\beq
\pi_k(Y^{p,q})=\pi_k(S^2)^2\hsp k>2.
\eeq

For example the third homotopy group is
\beq
\pi_3(Y^{p,q})=\Z^2. \label{p3}
\eeq
If one considers $Y^{p,q}$ to be a bundle over the $S^3$ with $S^2$ fibers, then one can visualise 
the two generators of $\Z^2$.  A generator of the first $\Z$ is just the map to a constant section of the bundle.  
One may act on such a section by an $SO(3)$ rotation of the $S^2$ at each point on the $S^3$.  
Such rotations are not necessarily connected to the identity, instead they are classified homotopically 
by maps from the $S^3$ to $SO(3)$.  The space of homotopy classes of such maps is $\Z$.  
After acting on a section with such a map one obtains a global section which is not continuously 
connected to the constant section.  The difference between such a section and the constant section 
is the second $\Z$ factor in (\ref{p3}).  

One may change the coordinates of the $S^2$ so that any global section is the north pole.  
In this case the global section becomes a constant section in the new coordinates.  
Such coordinates therefore are not homotopic to the original coordinates.  
They describe a homotopically distinct trivialization of the $S^2$-bundle.  
As such global sections are classified by $\Z$, there is a one parameter family of trivializations.  
The $\pi_3$'s of these trivializations are related by $T$ transforms in the $SL(2,\Z)$ automorphism group of $\Z^2$.  
These are large diffeomorphisms of $Y^{p,q}$.  
By fixing a trivialization of $Y^{p,q}$, we have chosen a particular trivialization of the $S^2$ bundle.  
However we may act on the fibers by such a large diffeomorphism to obtain 
any of the other trivializations.  For certain applications, one trivialization may be more 
desirable than another, for example a given brane can wrap a constant section in only one trivialization.  
In particular, it may be that there is a trivialization 
such that the calibrated cycles are constant sections or at least lie in only 
the first $\Z$ of $\pi_3(Y^{p,q})$.

\section{Helpful relations involving $y_1$, $y_2$ and $y_3$}
\label{AppC}

In this appendix we collect all helpful relations between the roots $y_i$
\beq
\frac{1}{6 \ell} \left( 1 -\frac{1}{y_1} \right) = \frac{p+q}{2}, \qquad
\frac{1}{6 \ell} \left( 1 -\frac{1}{y_2} \right) = -\frac{p-q}{2}
\eeq
and 
\beq
\frac{1-y_3}{1-y_1} = 1 + \frac{3 y_1}{y_2-y_1}, \qquad
\frac{1-y_3}{1-y_2} = 1 - \frac{3 y_2}{y_2-y_1} .
\eeq

\section{The $k=1$ case}
\label{AppD}

For $k=1$ the condition (\ref{eq:q2uw})  implies that $(\theta=\frac{\pi}{2},y=y_1)$ or $y=y_2$.
In the latter case $\theta$ remains unspecified. This is in contrast to the $k>1$ case,
where we found three isolated solutions $\theta=0$, $\frac{\pi}{2}$ and $\pi$.
We therefore cannot represent the integral as a sum of four Dirac surface contributions.
However, there is an infinitesimal deformation of (\ref{eq:q2uw}) that  brings the $k=1$
embedding to the form depicted on Figure \ref{Pict3}
\beq
u_2 \overline{w}_1 = \left| \frac{u_2}{u_1} \right|^\delta \overline{u}_1 \frac{w_2^{2}}{|w_2|},
\eeq
where $\delta$ is a small positive parameter. For $\delta=0$ we recover the original condition (\ref{eq:q2uw})
for $k=1$.  Obviously, for infinitesimally small $\delta$ the flux calculation
should not be different from the $\delta=0$ result.
Moreover, since $\delta<1$ the deformation is well-defined both near $u_1=0$ and $u_2=0$.
Finally, for $y=y_2$ (or $v_2=0$) we now have isolated solutions $\theta=0$, $\frac{\pi}{2}$ and $\pi$
exactly as in the $k>1$ case, while for $y=y_1$ (or $v_1=0$) there  
is still a single solution near  $\theta=\frac{\pi}{2}$.
The rest of the calculation proceeds exactly as in the $k>1$ case.

\bibliographystyle{unsrt}

\end{document}